\pdfoutput=1
\documentclass[amsmath,showpacs,superscriptaddress,floatfix,amssymb,onecolumn]{revtex4-1}
\usepackage[utf8]{inputenc}
\usepackage[english]{babel}
\usepackage{graphicx,float,dcolumn,bm,subfigure,epsf,natbib}
\begin{document}

\title{Suppression of chaos through coupling to an external chaotic system}
\author{Sudhanshu Shekhar Chaurasia\footnote{email : sudhanshushekhar@iisermohali.ac.in}}
	\author{Sudeshna Sinha\footnote{email : sudeshna@iisermohali.ac.in}}
	\affiliation{Indian Institute of Science Education and Research (IISER) Mohali, Knowledge City, SAS Nagar, Sector $81$, Manauli PO $140$ $306$, Punjab, India}
	
	{\footnotetext{ \protect\vspace*{2.5cm}} }

	\begin{abstract}

We explore the behaviour of an ensemble of chaotic oscillators coupled only to an external chaotic system, whose intrinsic dynamics may be similar or dissimilar to the group. Counter-intuitively, we find that a dissimilar external system manages to suppress the intrinsic chaos of the oscillators to fixed point dynamics, at sufficiently high coupling strengths. So, while synchronization is induced readily by coupling to an identical external system, control to fixed states is achieved only if the external system is dissimilar. We quantify the efficacy of control by estimating the fraction of random initial states that go to fixed points, a measure analogous to basin stability. Lastly, we indicate the generality of this phenomenon by demonstrating suppression of chaotic oscillations by coupling to a common hyper-chaotic system. These results then indicate the easy controllability of chaotic oscillators by an external chaotic system, thereby suggesting a potent method that may help design control strategies.
		
	\end{abstract}
	
	\maketitle
		
	\section{Introduction}

        The emergence of steady states has been observed in many complex systems, such as chemical reactions \citep{steady1,steady2} and biological oscillators \citep{steady3,steady4,steady5,neuron}. Such fixed dynamics may be the desired target in certain cases, for instance in laser systems \citep{laser1,laser2,laser3}, where it leads to stabilization. On the other hand, the suppression of oscillations can signal pathology, such as in neuronal disorders like Alzheimer or Parkinson’s disease \citep{alzheimer1,alzheimer2,alzheimer3}. In this context too it is important to know in which situations steady states arise. Further, from the point of view of human-engineered systems it is important to find control methods that can effectively and efficiently tame chaotic dynamics \citep{control,eng1,eng2,eng3,eng4}. For all these reasons, there has been considerable sustained research on suppression of chaotic oscillations in nonlinear systems over the years.
        
\begin{figure}[H]
		\centering
		\includegraphics[width=0.25\linewidth]{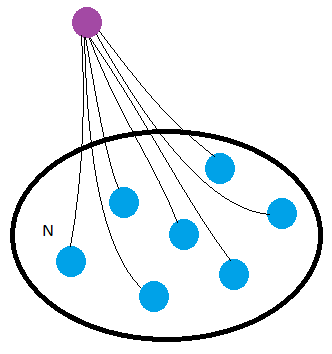}
		\caption{Schematic of a group of $N$ oscillators coupled to an external oscillator.}
		\label{schematic}
		\end{figure}

        In this work we explore the behaviour of an ensemble of chaotic oscillators coupled only to an external chaotic system. So there is no direct coupling amongst the oscillators, and the interaction is mediated by coupling to the common external system  \citep{resmi}. So this external system can be thought of as a pacemaker of the group of oscillators (cf. Fig.~\ref{schematic} for a schematic). Note that the intrinsic dynamics of the external system can be {\em identical} to the group, or it can be an entirely {\em different} type of dynamical system.
        
        
        

		Specifically, we first consider the example of $N$ R{\"o}ssler oscillators in a group, labelled by index $i= 1, \dots N$, with dynamics given by: 

		\begin{eqnarray}
		\label{rossler_group}
		\dot{x_i}&=&-(\omega + \delta (x_i^2+y_i^2)) \ y_i - z_i + \varepsilon \ ( x_{ext} - x_i ) \nonumber \\
		\dot{y_i}&=&(\omega + \delta (x_i^2+y_i^2)) \ x_i + a \ y_i \\
		\dot{z_i}&=&b + z_i (x_i-c) \nonumber
		\end{eqnarray}
		
where $x_{ext}$ is a dynamical variable of the common external system to which the group is coupled diffusively.
The strength of coupling is given by $\varepsilon$.

When the external oscillator is also a R{\"o}ssler oscillator, its governing equations are given by:

		\begin{eqnarray}
		\label{rossler_external}
		\dot{x}_{ext}&=&-(\omega + \delta (x_{ext}^2+y_{ext}^2)) \ y_{ext} - z_{ext} + \ \frac{\varepsilon}{N} \sum_{j=1}^N \ ( x_j - x_{ext} ) \nonumber \\
		\dot{y}_{ext}&=&(\omega + \delta (x_{ext}^2+y_{ext}^2)) \ x_{ext} + a \ y_{ext} \\
		\dot{z}_{ext}&=&b +z_{ext} (x_{ext}-c) \nonumber
		\end{eqnarray}
		
		When the external oscillator is distinct from the group, for instance a Lorenz system, its dynamical equations are given by:

		\begin{eqnarray}
		\label{lorenz}
		\dot{x}_{ext}&=&\sigma \ (y_{ext} - x_{ext}) \ + \ \frac{\varepsilon}{N}\sum_{j=1}^N \ ( x_j - x_{ext} ) \nonumber \\
		\dot{y}_{ext}&=&x_{ext} \ (r - z_{ext}) \ - \ y_{ext} \\
		\dot{z}_{ext}&=&x_{ext} \ y_{ext} \ - \ \beta \ z_{ext} \nonumber
		\end{eqnarray}
		
		Parameters $\sigma$, $\beta$, $r$ in the Lorenz system and parameters $a$, $b$, $c$, $\omega$, $\delta$ in the R{\"o}ssler oscillator, regulate the nature of the uncoupled dynamics, which can range from fixed points to chaos. In the sections below, we will present the spatiotemporal patterns arising in two distinct situations of interest: (a) the group of oscillators and the external oscillator are of identical type, and (b) the external chaotic system is distinct from the group of oscillators, and may even be hyper-chaotic.

	\section{Emergent Controlled Dynamics}

\begin{figure}[H]
			\centering

			\includegraphics[width=0.485\linewidth]{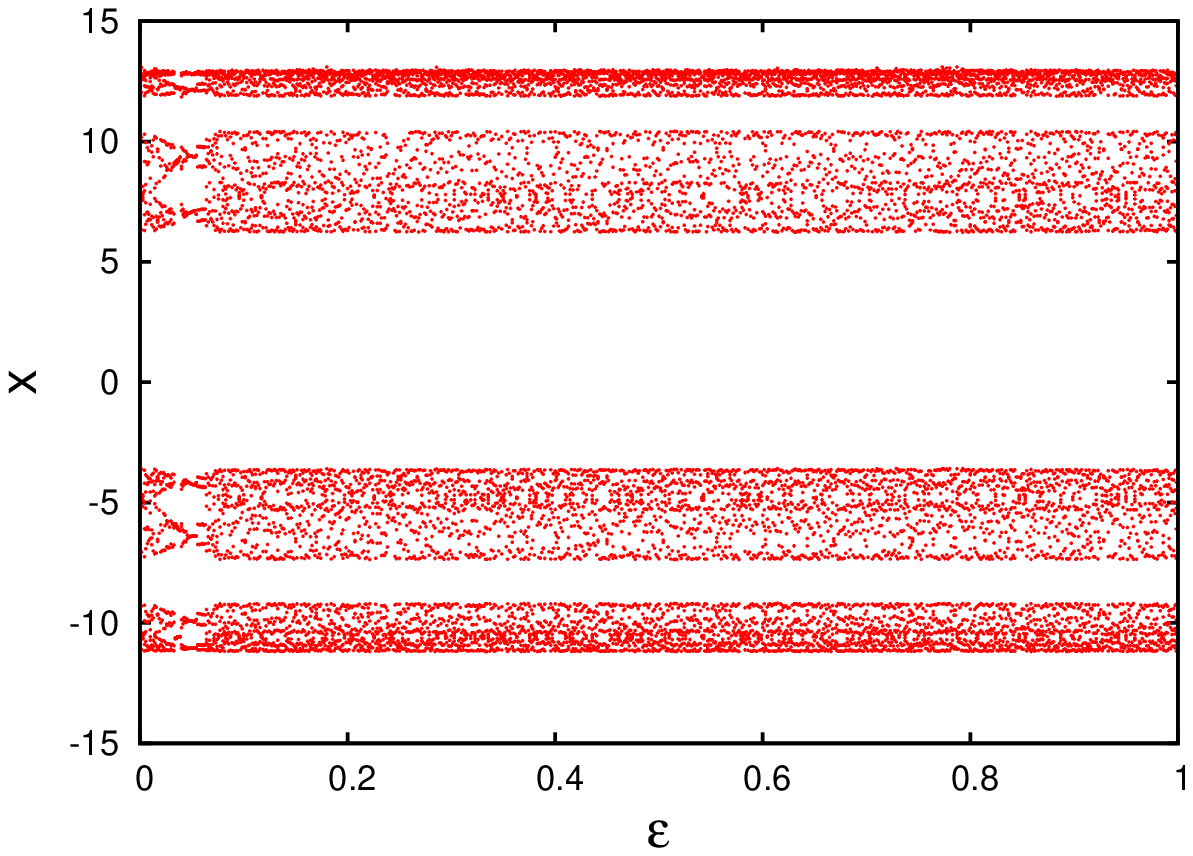}
			\includegraphics[width=0.485\linewidth]{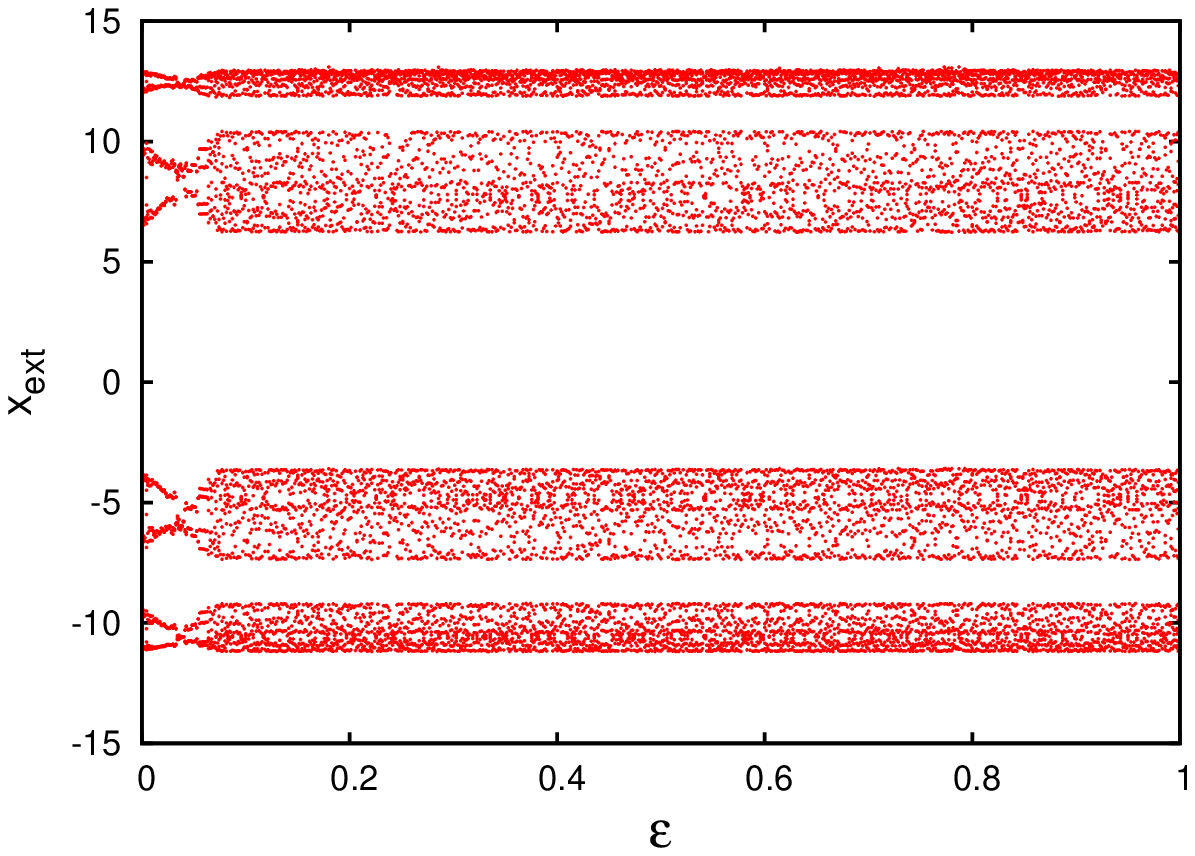}

(a)

			\includegraphics[width=0.485\linewidth]{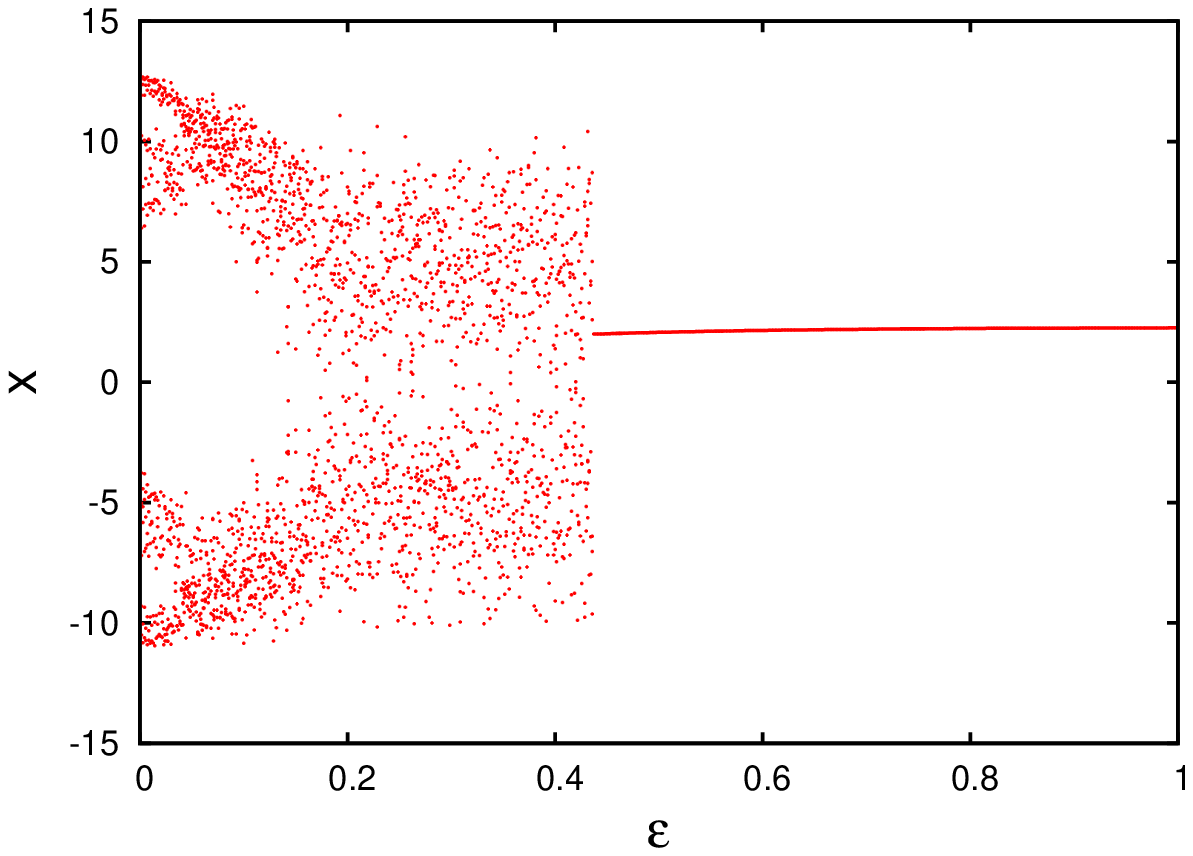}
			\includegraphics[width=0.485\linewidth]{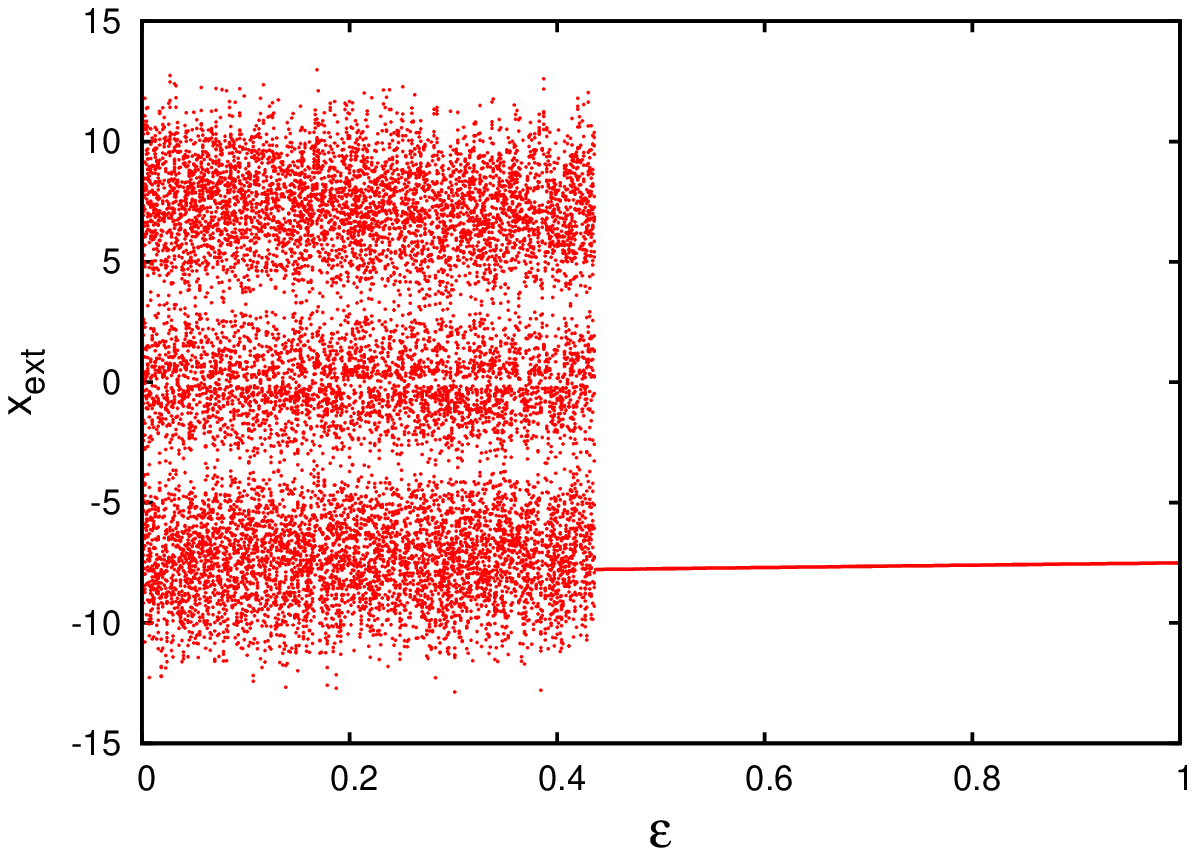}
			
(b)
			\caption{Bifurcation diagrams, with respect to the coupling strength  $\varepsilon$, of one representative oscillator in the group (left) and an external oscillator (right). Here the group consists of chaotic R{\"o}ssler oscillators with parameters $\omega=0.41$, $\delta=0.0026$, $a=0.15$, $b=0.4$ and $c=8.4$  in Eqn.~\ref{rossler_group}, and the external oscillator is: (a) a chaotic R{\"o}ssler oscillator with parameters  $\omega=0.41$, $\delta=0.0026$, $a=0.15$, $b=0.4$ and $c=8.4$ in Eqn.~\ref{rossler_external}, and (b) a chaotic Lorenz system with parameters $\sigma=10.0$, $\beta=8.0/3.0$ and $r=25.0$ in Eqn.~\ref{lorenz}. 
In all the diagrams (including ones below) we display the $x$ variable on the Poincare section of the phase curves of the oscillators at $y = y_{\text{mid}}$, where $y_{\text{mid}}$ is the mid-point of the span of attractors along the $y$-axis.}
			\label{hub_chaotic_endnode_chaotic}
		\end{figure}

Fig. \ref{hub_chaotic_endnode_chaotic} shows the bifurcation diagrams of the illustrative cases of an ensemble of chaotic R{\"o}ssler oscillators coupled to (a) a chaotic external system that is identical (namely another R{\"o}ssler oscillator) and (b) a chaotic external system that is dissimilar (namely, a Lorenz system). We find that a group of chaotic oscillators can be controlled to fixed points by the external {\em dissimilar} chaotic oscillator, when coupling is stronger than a critical value. However, when the chaotic oscillators are coupled to an external chaotic system of an {\em identical} type (namely all are R{\"o}ssler oscillators), none of the oscillators are controlled to fixed states, even for strong coupling.

\begin{figure}[H]
			\centering
%

			\includegraphics[width=0.75\linewidth]{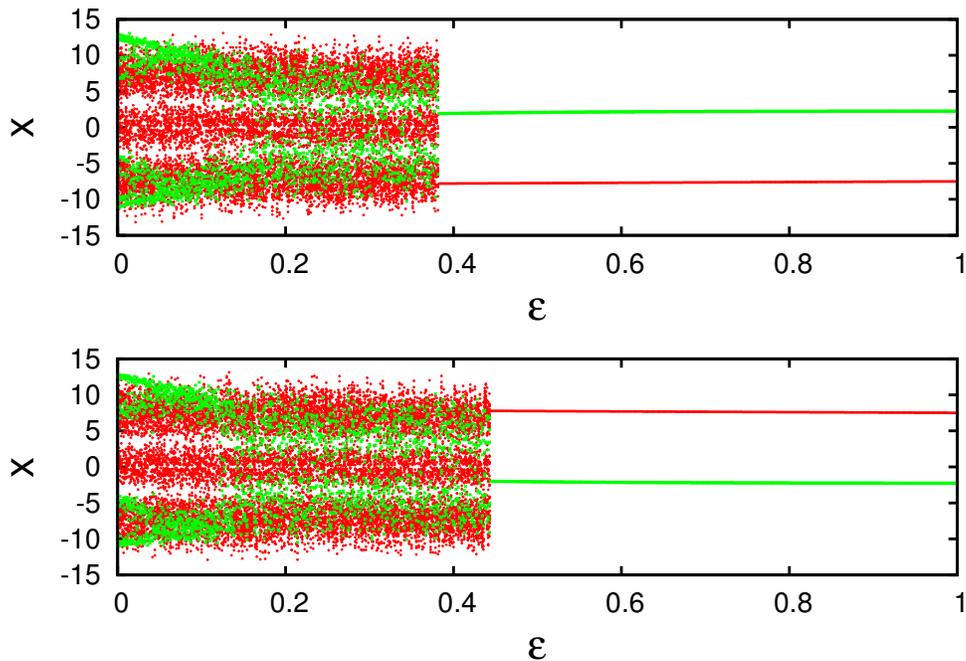}
			

			\caption{Bifurcation diagrams, with respect to the coupling strength  $\varepsilon$, of one representative oscillator in the group (green) and an external oscillator (red), for the cases where the group of oscillators is comprised of chaotic R{\"o}ssler oscillators with parameters $\omega=0.41$, $\delta=0.0026$, $a=0.15$, $b=0.4$ and $c=8.4$  in Eqn.~\ref{rossler_group} and the external oscillator is a 
chaotic Lorenz system with parameters $\sigma=10.0$, $\beta=8.0/3.0$ and $r=25.0$ in Eqn.~\ref{lorenz}.} 
			\label{two_fp}
		\end{figure}

When a group of chaotic R{\"o}ssler oscillators is coupled on a common external chaotic Lorenz system, we find that there exists two steady states, as illustrated in Fig.~\ref{two_fp}. Depending on initial conditions, the system can go to either of the steady states.  Linear stability analysis, via eigenvalues of the Jacobian, also corroborates the stabilization of the fixed points seen in the bifurcation diagrams (see Appendix for details).

	\section{Synchronization}
		
        We study the advent of synchronization in the group of oscillators, as a function of the coupling strength, for the case of identical and distinct external systems. Our focus is to ascertain what kind of external system facilitates synchrony, and which ones lead to control to steady states.

		\begin{figure}[H]
			\centering
			

			\includegraphics[width=0.55\linewidth]{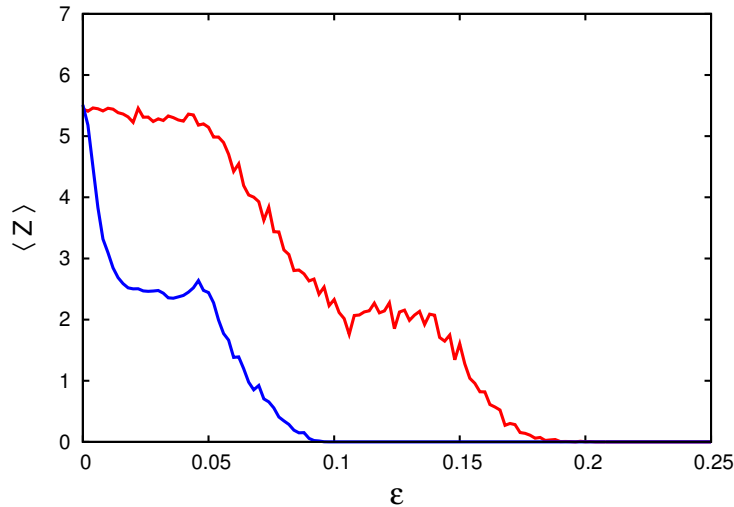}
			

			\caption{Synchronization error $\langle Z \rangle$ of the chaotic R{\"o}ssler oscillators in the group with parameters $\omega=0.41$, $\delta=0.0026$, $a=0.15$, $b=0.4$ and $c=8.4$  in Eqn.~\ref{rossler_group}, averaged over $100$ different initial conditions, with respect to coupling strength  $\varepsilon$, for the case where the common external system is an identical chaotic R{\"o}ssler oscillator (blue) with parameters $\omega=0.41$, $\delta=0.0026$, $a=0.15$, $b=0.4$ and $c=8.4$  in Eqn.~\ref{rossler_external} and a chaotic Lorenz attractor (red)  with parameters $\sigma=10.0$, $\beta=8.0/3.0$ and $r=25.0$ in Eqn.~\ref{lorenz}.}
			\label{sync_err_lorenz}
		\end{figure}

We calculate the synchronization error of the group of oscillators, averaged over time $T$, given by 
		$$Z = \frac{1}{T}\sum_t \sqrt{\frac{1}{N}\sum_{i=1}^N (x_i-\bar{x})^2}$$
		where $\bar{x}$ is the average value of the $x$ variable of oscillators $i = 1, \dots N$, at an instant of time. Further we average $Z$ over different initial states to obtain an ensemble averaged synchronization error $\langle Z \rangle$.

We display this average synchronization error in Fig.~\ref{sync_err_lorenz}, from where it is clearly evident that as coupling strength increases, the group of uncoupled oscillators coupled to a common external chaotic oscillator get synchronized. This trend holds for both identical and distinct external oscillators, suggesting that coupling to an external chaotic oscillator of wide-ranging dynamical types can induce synchronization. The critical coupling at which synchronization occurs is higher when the common external oscillator is distinct from the group. 

Interestingly, as coupling strength increases further, the oscillators are controlled to steady states, when the external oscillator is {\em dissimilar}. So an identical common external oscillator induces synchronization at weaker coupling strengths than a dissimilar external oscillator, but control to steady states occurs {\em only} when the external oscillator is distinct from the group.
	
\section{Basin Stability of the Spatiotemporal Fixed Point}

We now quantify the efficacy of control to steady states by finding the fraction $BS_{fixed}$ of initial states that are attracted to fixed points, starting from generic random initial conditions. This measure is analogous to recently used measures of {\em basin stability} \citep{menck}, and indicates the size of the basin of attraction for a spatiotemporal fixed point state.  $BS_{fixed} \sim 1$ suggests that the fixed point state is globally attracting, while $BS_{fixed} \sim 0$ indicates that almost no initial states evolve to stable fixed states. 

\begin{figure}[H]
			\centering
			\includegraphics[width=0.55\linewidth]{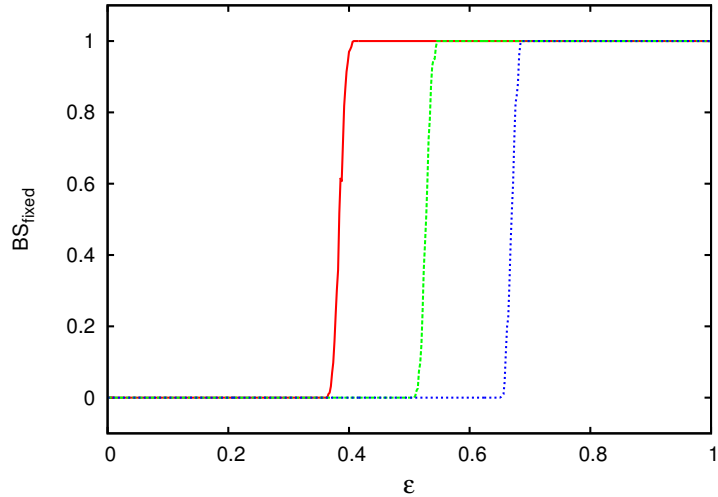}
						\caption{Dependence of the fraction of initial states $BS_{fixed}$ attracted to the fixed point state, on the coupling strength $\varepsilon$, for a group of chaotic R{\"o}ssler oscillators with parameters $\omega=0.41$, $\delta=0.0026$, $a=0.15$, $b=0.4$ and $c=8.4$  in Eqn.~\ref{rossler_group}, coupled to a common external chaotic Lorenz system with parameters $\sigma=10.0$, $\beta=8.0/3.0$ and $3$ different values of parameter $r$ in Eqn.~\ref{lorenz}: $24.7$ (red), $25.0$ (green) and $25.3$ (blue). Note that there is no dependence of $BS_{fixed}$ on the number of oscillators $N$ in the group.}
			\label{basin1}
		\end{figure}

\begin{figure}[H]
			\centering
			\includegraphics[width=0.55\linewidth]{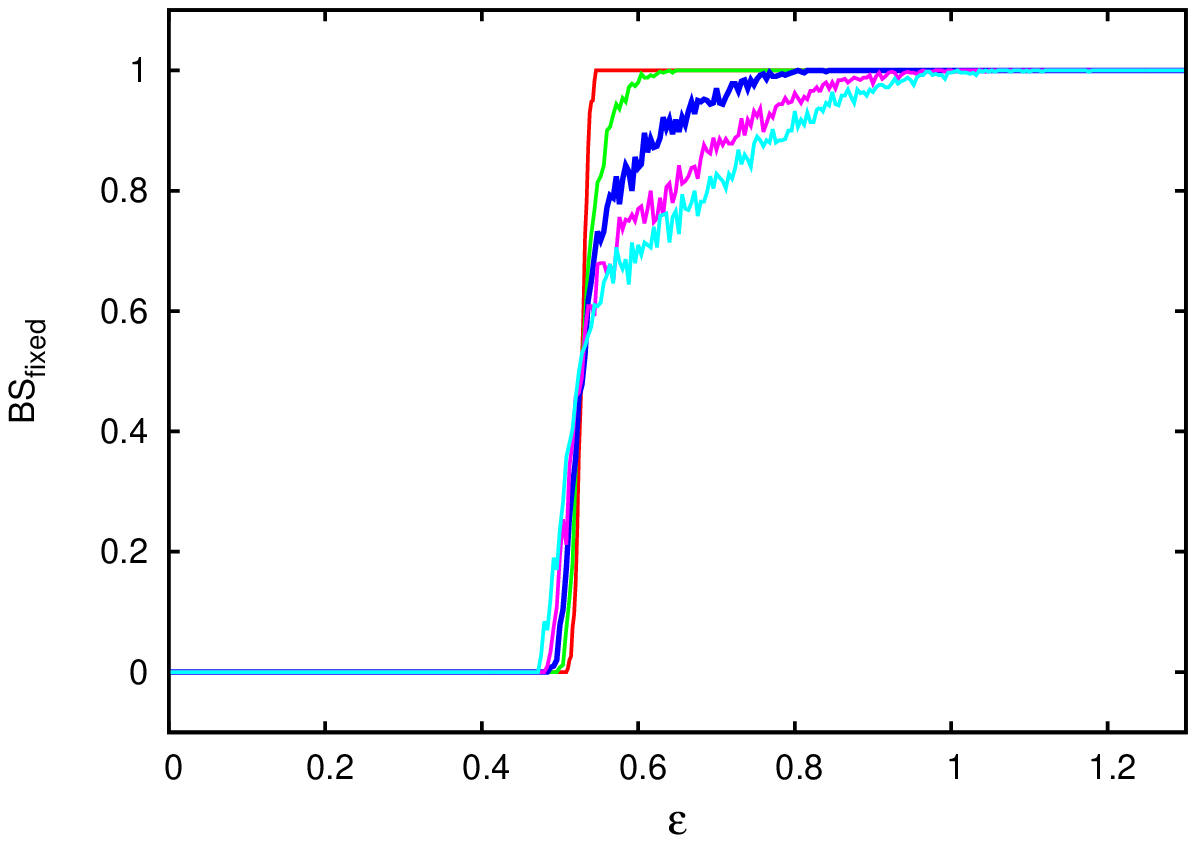}

(a)

			\includegraphics[width=0.55\linewidth]{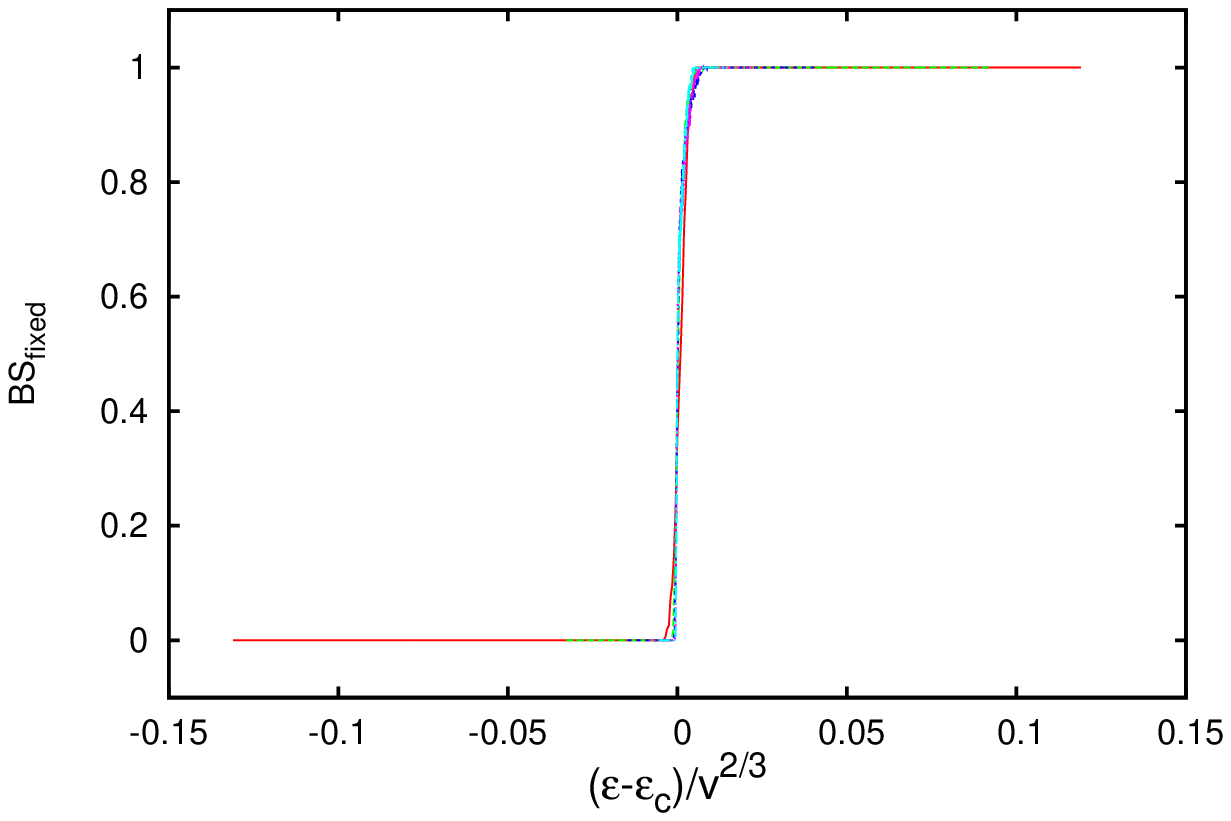}

(b)



						\caption{Dependence of the fraction of initial states $BS_{fixed}$ attracted to the fixed point state, on the coupling strength $\varepsilon$, for a group of chaotic R{\"o}ssler oscillators with parameters $\omega=0.41$, $\delta=0.0026$, $a=0.15$, $b=0.4$ and $c=8.4$  in Eqn.~\ref{rossler_group}, coupled to a common external chaotic Lorenz system with parameters $\sigma=10.0$, $\beta=8.0/3.0$ and $r=25.0$ in Eqn.~\ref{lorenz}. In (a) the $5$ curves are obtained from  initial states randomly distributed in a box of linear size $l=2, 4, 6, 8, 10$ in the $x$, $y$ and $z$ coordinates, and in (b) we show data collapse by appropriate scaling, with $v = l^3$, and $\varepsilon_c=0.524$.}
			\label{basin2}
		\end{figure}

We  show the dependence of this fraction $BS_{fixed}$ in Figs.~\ref{basin1}-\ref{basin2} for a group of chaotic R{\"o}ssler oscillators coupled to an external chaotic Lorenz system as coupling strength is varied. It is evident that there is a sharp transition to complete control, where the spatiotemporal fixed point state is globally attracting, at sufficiently strong coupling. Namely, there is a critical coupling strength beyond which the intrinsic chaos of the oscillators is suppressed to fixed points, over a very large basin of initial states. 

Further, we show this transition for different external Lorenz systems. In particular, we vary parameter $r$ in Eqn.~\ref{lorenz}. This changes the lyapunov exponent of the intrinsic dynamics of the external system, which increases linearly with $r$. It is clearly observed that as the lyapunov exponent of the intrinsic dynamics of the external system increases, the transition shifts to higher coupling strengths. Namely, it takes stronger coupling to yield control to steady states as the common (dissimilar) external system gets more chaotic.

\section{Control to Steady States via an External Hyper-chaotic Oscillator}

We have checked the generality of the results by considering a more stringent case of a group of chaotic oscillators coupled to an external hyper-chaotic oscillator, given by:
		\begin{eqnarray}
		\label{eqn:hyperchaos}
		\dot{x}_{ext}&=&(k'-2)x_{ext} - y_{ext} - G(x_{ext}-z_{ext}) + \frac{\varepsilon}{N}\sum_{j=1}^N (x_j - x_{ext}) \nonumber \\
		\dot{y}_{ext}&=&(k'-1)x_{ext} - y_{ext} \\
		\dot{z}_{ext}&=& -w_{ext} + G(x_{ext}-z_{ext}) \nonumber \\
		\dot{w}_{ext}&=& \beta' z_{ext} \nonumber \\ \nonumber \\
		\text{where} \ \ \ \ \ G(u) &=& \frac{1}{2}b' \{ |u-1| + (u-1) \} \nonumber
		\end{eqnarray}
		
		where $k'$, $\beta'$, $b'$ are the parameters determining the dynamics of the oscillator.

We have coupled one variable (specifically, $x_{ext}$) of the hyper-chaotic external oscillator with one variable (specifically, $x$) of the group of chaotic R{\"o}ssler oscillators.
		
		\begin{figure}[H]
			\centering
			\includegraphics[width=0.485\linewidth]{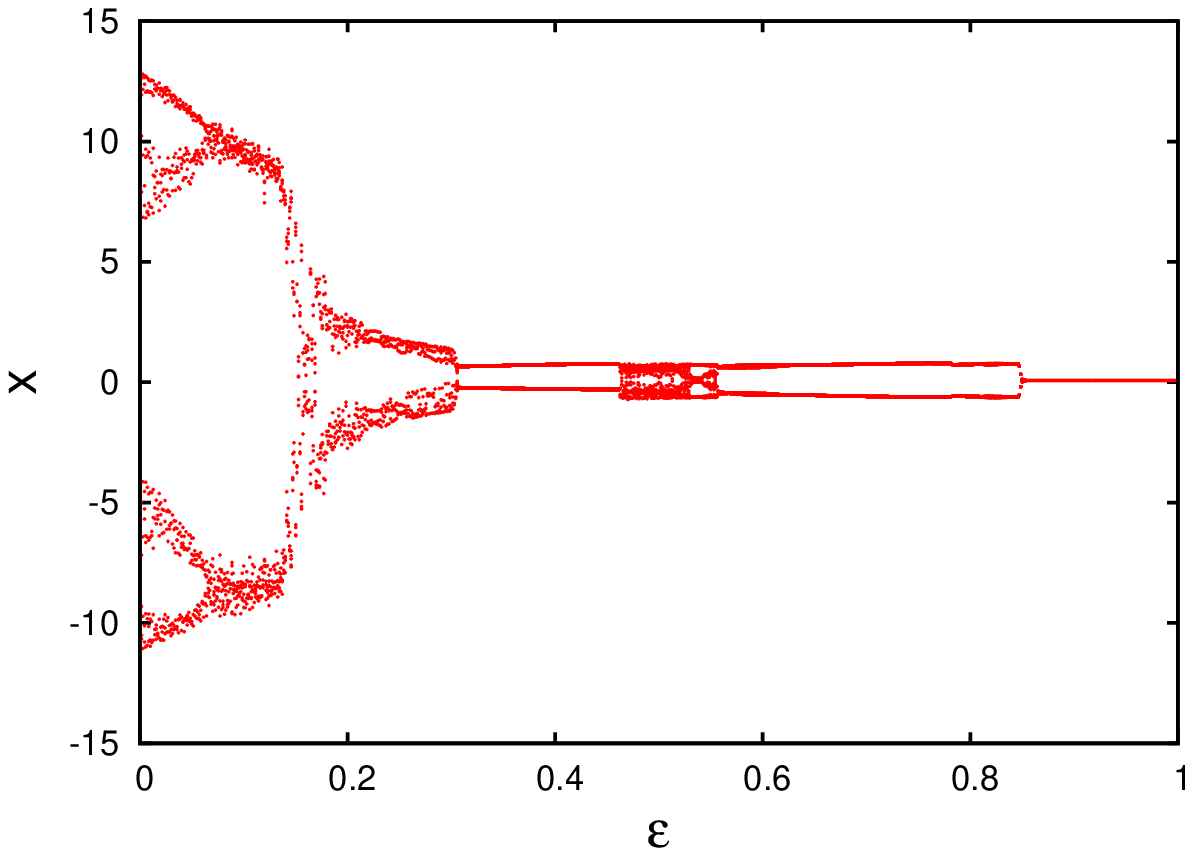}
			\includegraphics[width=0.485\linewidth]{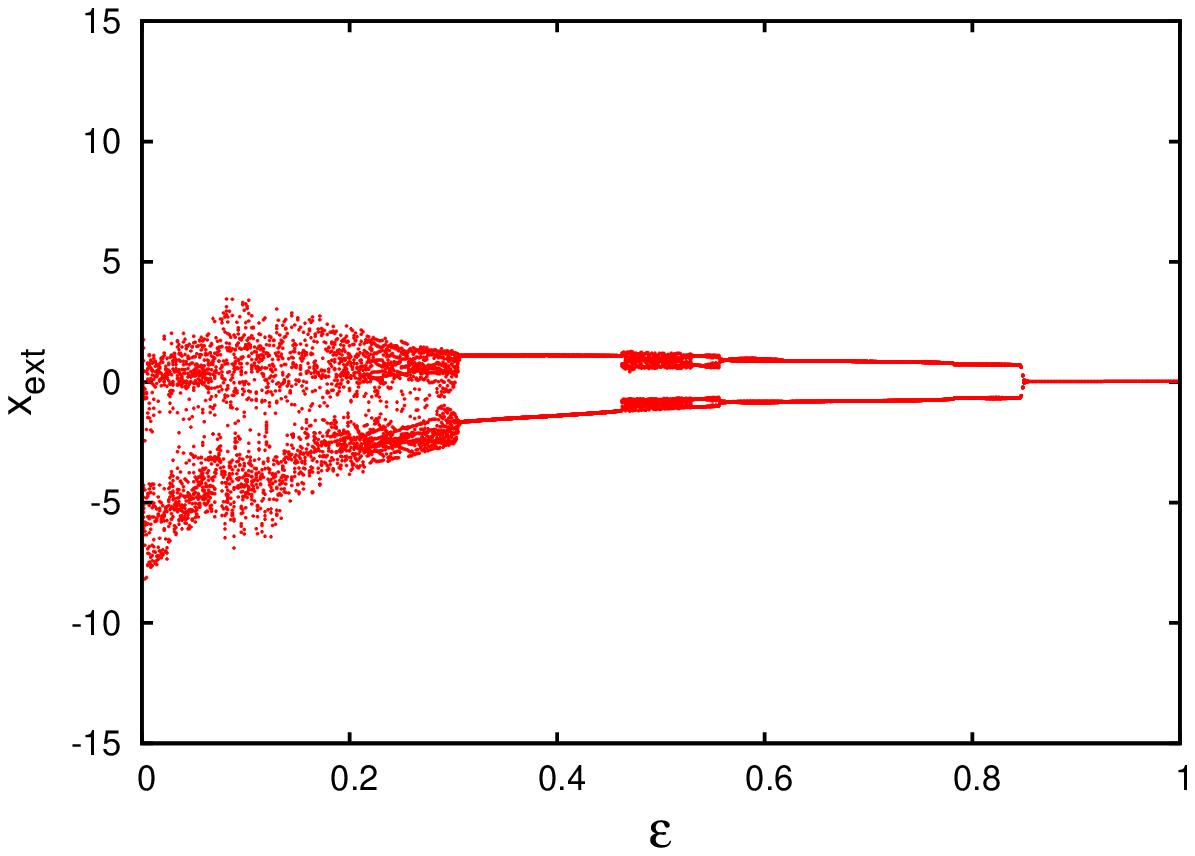}

			\caption{Bifurcation diagram, with respect to the coupling strength  $\varepsilon$, of one representative oscillator of the group (left) and  the external system (right), when  the external system is hyper-chaotic with parameters $k'=3.85$, $\beta'=18.0$ and $b'=88.0$ in Eqn.~\ref{eqn:hyperchaos}, and the group consists of chaotic R{\"o}ssler oscillators with parameters $\omega=0.41$, $\delta=0.0026$, $a=0.15$, $b=0.4$ and $c=8.4$ in Eqn.~\ref{rossler_group}.}
			\label{fig:hub_hyperchaotic_endnode_chaotic}
		\end{figure}
		
\begin{figure}[H]
			\centering
			\includegraphics[width=0.6\linewidth]{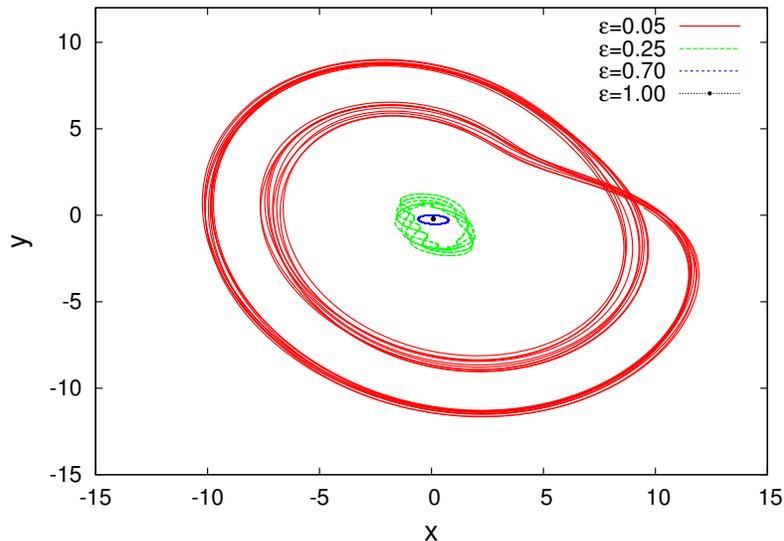}
			\caption{Phase portrait of a representative R{\"o}ssler oscillator from the group  with parameters $\omega=0.41$, $\delta=0.0026$, $a=0.15$, $b=0.4$ and $c=8.4$ in Eqn.~\ref{rossler_group}, coupled to a common external hyper-chaotic oscillator with parameters $k'=3.85$, $\beta'=18.0$ and $b'=88.0$ in Eqn.~\ref{eqn:hyperchaos}, at different coupling strengths  $\varepsilon$.}
			\label{fig:Phase_Space_Rossler_whenHubHyperchaotic}
		\end{figure}

		Interestingly, we again find that the intrinsically chaotic R{\"o}ssler oscillators go to fixed points, when coupled to a common external hyper-chaotic oscillator, for sufficiently strong coupling. Fig.~\ref{fig:Phase_Space_Rossler_whenHubHyperchaotic} shows the phase portrait of the R{\"o}ssler oscillators at different coupling strengths. It is apparent that the group of oscillators become regular when coupling strengths are sufficiently high. 
		
\begin{figure}[H]
			\centering
			\includegraphics[width=0.6\linewidth]{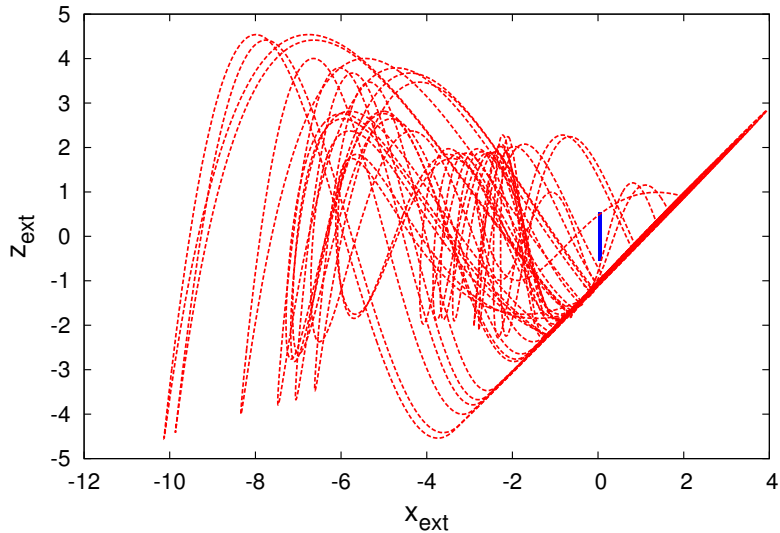}
			\caption{Phase portrait of the external hyper-chaotic oscillator with parameters $k'=3.85$, $\beta'=18.0$ and $b'=88.0$ in Eqn.~\ref{eqn:hyperchaos}, for the case of an uncoupled external oscillator (red) and an oscillator with coupling to a group of chaotic R{\"o}ssler oscillator with parameters $\omega=0.41$, $\delta=0.0026$, $a=0.15$, $b=0.4$ and $c=8.4$ in Eqn.~\ref{rossler_group}, with coupling strength $\varepsilon=1.0$ (blue).}
			\label{fig:Phase_Space_Hub_Hyperchaotic}
		\end{figure}

		Fig.~\ref{fig:Phase_Space_Hub_Hyperchaotic} shows the phase portrait for the external hyper chaotic oscillator. Again it is clear that at high coupling strengths, the dynamics of the hyper-chaotic system becomes regular. Further the size of the emergent external attractor is very small, though not a fixed point.
		
Note that when coupling strength $\varepsilon > \varepsilon_c$, the $(x_{ext}-z_{ext}-1)$ term in Eqn.~\ref{eqn:hyperchaos} becomes less than zero, implying that $G(x_{ext}-z_{ext})$ is always zero. So the dynamical equations for $\dot{x}_{ext}$ and $\dot{z}_{ext}$ becomes uncoupled, yielding two independent sub-sets of equations, with one coupled sub-set comprising of $\dot{x}_{ext}$ and $\dot{y}_{ext}$, and another coupled sub-set comprising of $\dot{z}_{ext}$ and $\dot{w}_{ext}$.
		

	\section{Conclusions}

We investigated the behaviour of an ensemble of uncoupled chaotic oscillators coupled diffusively to an external chaotic system. The common external system may be similar or dissimilar to the group. We explored all possible scenarios, with the intrinsic dynamics of the external system ranging from chaotic to hyper-chaotic. Counter-intuitively, we found that an external system manages to successfully steer a group of chaotic oscillators on to steady states at sufficiently high coupling strengths when it is {\em dissimilar} to the group, rather than identical. So while the group of oscillators coupled to an identical external system synchronizes readily, surprisingly enough, control to fixed states is achieved {\em only} if the external oscillator is dissimilar. We indicate the generality of this phenomenon by demonstrating the suppression of chaotic oscillations by coupling to an external hyper-chaotic system. 

Further, for the case of coupling to a non-identical external system, we quantified the efficacy of control by estimating the fraction of generic random initial states where the intrinsic chaos of the oscillators is suppressed to fixed points, a measure analogous to basin stability. We showed that there was a sharp transition to complete control, where the spatiotemporal fixed point is a global attractor, after a critical coupling strength.

In summary, our results demonstrate robust control of a group of chaotic oscillators to fixed points, by coupling to a dissimilar external chaotic system. We thus suggest a potent way to tame chaotic dynamics that may occur in natural systems, or may be used in design of control strategies in engineering contexts.
		
\bigskip		
\bigskip

	\section*{Appendix : Stability Analysis}
		
\begin{figure}[H]
			\centering
			\includegraphics[width=0.475\linewidth]{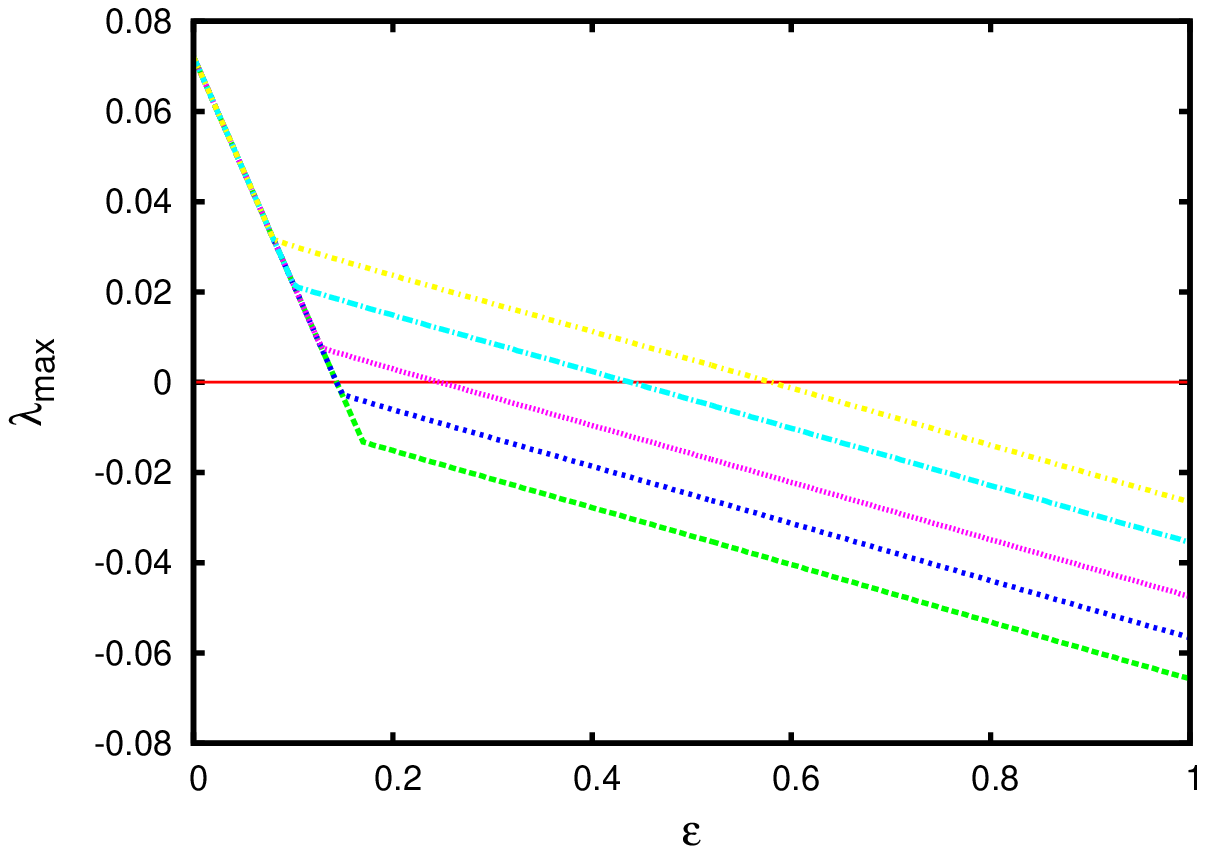}
%
%
			\includegraphics[width=0.475\linewidth]{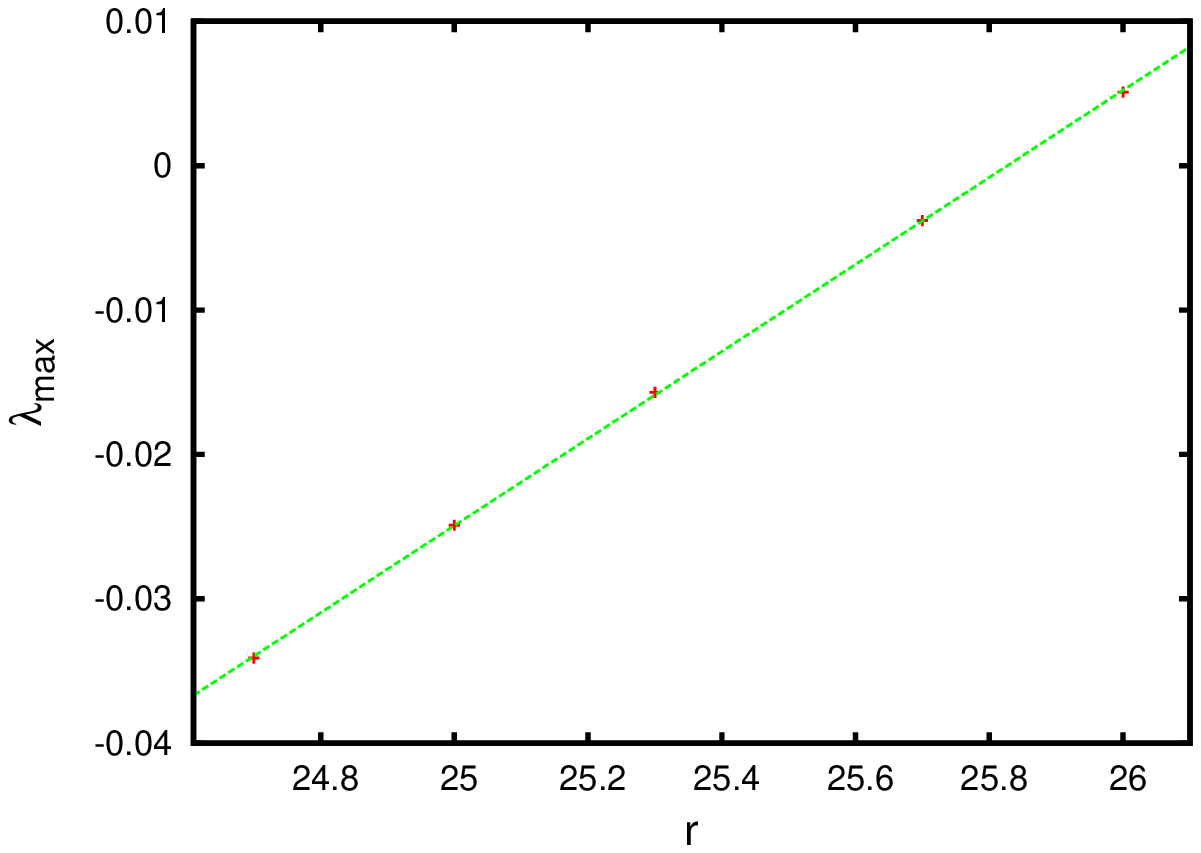}

\hspace{1.9in} (a) \hfill  (b) \hspace{1.5in}

			\caption{(a) Maximum real part $\lambda_{max}$ of the eigenvalues of the Jacobian at the fixed points, as a function of coupling strength $\varepsilon$, for the external Lorenz system having parameters $\sigma=10.0$, $\beta=8.0/3.0$ and $r=25.0$ and (bottom to top) $r = 24.7, 25.0, 25.3, 25.7, 26.0$ in Eqn.~\ref{lorenz} and group of chaotic R{\"o}ssler oscillator with parameters $\omega=0.41$, $\delta=0.0026$, $a=0.15$, $b=0.4$ and $c=8.4$ in Eqn.~\ref{rossler_group}. (b)  Dependence of $\lambda_{max}$ on parameter $r$ of the external Lorenz system, at fixed coupling strength ($\varepsilon=0.5$ here). }
			\label{fig:jac_hub_lorenz_end_rossler}
		\end{figure}

		We investigate the linear stability of the steady state obtained when a group of intrinsically chaotic R{\"o}ssler oscillators is coupled to a common external intrinsically chaotic Lorenz system, via the eigenvalues of the Jacobian matrix evaluated at those fixed points. Specifically, the Jacobian for $N$ number of oscillators and an external oscillator is given by $3(N+1) \times 3(N+1)$ matrix. Calculating each term using Eqns.~(\ref{rossler_group}) and (\ref{lorenz}), we obtain the Jacobian matrix to be:
		
		$$J=\begin{pmatrix}
		-\sigma - \varepsilon & \sigma & 0    & \frac{\varepsilon}{N}                          & 0                                      & 0         & . & . & . & .\\
		r-z_0                      & -1  & -x_0 & 0                                               & 0                                      & 0         & . & . & . & .\\
		y_0                          & x_0 & -\beta & 0                                               & 0                                      & 0         & . & . & . & .\\
		\varepsilon        & 0   & 0    & -2\delta x_1 y_1 - \varepsilon & -\omega - \delta (x_1^2+3y_1^2) & -1         & . & . & . & .\\
		0                            & 0   & 0    & \omega + \delta (3x_1^2+y_1^2)           & 2\delta x_1 y_1+a             & 0         & . & . & . & .\\
		0                            & 0   & 0    & z_1                                             & 0                                      & x_1 - c & . & . & . & .\\
		.                            & .   & .    & .                                               & .                                      & .         & . & . & . & .\\
		.                            & .   & .    & .                                               & .                                      & .         & . & . & . & .\\
		.                            & .   & .    & .                                               & .                                      & .         & . & . & . & .\\
		.                            & .   & .    & .                                               & .                                      & .         & . & . & . & .\\
		\end{pmatrix}$$
		$ $ \\
		At each coupling strength, there is a set of $3(N+1)$ eigenvalues of the Jacobian. We show the maximum real part $\lambda_{max}$ of the eigenvalues as a function of coupling strength  $\varepsilon$ in Fig.~\ref{fig:jac_hub_lorenz_end_rossler}a. Naturally, since the group of oscillators and the external oscillator are intrinsically chaotic, $\lambda_{max} > 0$ for $\varepsilon = 0$. However, it is clear that at a critical value of coupling $\varepsilon_c$ all eigen values are negative, indicating that all the oscillators in the group and the external system go to stable fixed points (cf. Fig.~\ref{fig:jac_hub_lorenz_end_rossler}a). Note that the  $\varepsilon_c$ obtained through linear stability analysis is smaller than that observed from generic random initial states, as displayed in the bifurcation plots. So we undertook additional numerical simulations from initial states sufficiently close to the fixed point solutions and verified that for such close-by initial conditions the fixed point state is indeed stable at lower coupling strengths, in accordance with that seen in  Fig.~\ref{fig:jac_hub_lorenz_end_rossler}a.

Further, from Fig.~\ref{fig:jac_hub_lorenz_end_rossler}b it is clear that $\lambda_{max}$ increases linearly with increasing parameter $r$ in the external Lorenz system (cf. Eqn.~\ref{lorenz}). This supports the numerical observations that the {\em critical coupling strength $\varepsilon_c$ increases linearly with $r$}, as displayed in Fig.~\ref{basin1}.
		
\bigskip
\bigskip
\bigskip

		\bibliography{arxiv_paper.bib}
		
\end{document}